\documentclass[12pt]{article}
\usepackage{amsfonts}

\newcommand{\bi}{\begin{itemize}}
\newcommand{\ei}{\end{itemize}}
\newcommand{\ioverh}{\frac{i}{\hbar}}
\newcommand{\Tr}{\mathop{\rm Tr}\nolimits}
\newcommand{\al}{\alpha}
\newcommand{\be}{\begin{equation}}
\newcommand{\ee}{\end{equation}}
\newcommand{\ba}{\begin{eqnarray}}
\newcommand{\ea}{\end{eqnarray}}
\newcommand{\ban}{\begin{eqnarray*}}
\newcommand{\ean}{\end{eqnarray*}}
\newcommand{\eq}[1]{(\ref{#1})}
\newcommand{\eqs}[2]{(\ref{#1},~\ref{#2})}
\newcommand{\eqss}[3]{(\ref{#1},~\ref{#2},~\ref{#3})}
\newcommand{\Eq}{Eq.~\eq}
\newcommand{\Eqs}{Eqs.~\eqs}

\newcommand{\Sect}[1]{Sect.~\ref{#1}}
\newcommand{\ra}{\rangle}

\newcommand{\E}{{\cal E}}
\newcommand{\R}{{\cal R}}
\newcommand{\Sys}{{\cal S}}

\begin{document}

\title{Evolution of an open system \\
as a continuous measurement \\
of this system by its environment\thanks{Reported at 3d Sakharov
Conference on Physics, Moscow, June 2002}}

\author{Michael~B.~Mensky\\
{P.N.Lebedev Physics Institute, 117924 Moscow, Russia}}

\date{} 

\maketitle

\begin{abstract}
The restricted-path-integral (RPI) description of a continuous quantum
measurement is rederived starting from the description of an open
system by the Feynman-Vernon influence functional. For this end the
total evolution operator of the compound system consisting of the
open system and its environment is decomposed into the sum of partial
evolution operators. Accordingly, the influence functional of the open
system is decomposed into the integral of partial influence
functionals (PIF). If the partial evolution operators or PIF are
chosen in such a way that they decohere (do not interfere with each
other), then the formalism of RPI effectively arises. The evolution of
the open system may then be interpreted as a continuous measurement of
this system by its environment. This is possible if the environment is
macroscopic or mesoscopic.
\end{abstract}

\section{Introduction}

When a quantum system is measured, its state is unavoidably changed.
This change may be described as decoherence (see \cite{Zeh-bk96} and
references therein). Continuous measuring a system leads to its
gradual decoherence. This type of evolution seems to be general, so
that the evolutions of a wide class of open systems may be described as
continuous fuzzy measurements of these systems by their environments
\cite{Men00bk}. In the present paper we shall formulate this statement
in more detail and present additional arguments supporting it.

Usually the evolution of a continuously measured (open, decohering)
system is described non-selectively, i.e. without taking into account
the concrete readout of the measurement (the concrete state of the
environment). In this case the state of the system is presented by a
density matrix which in Markovian approximation satisfies the
time-differential equation called master equation. The most general
form of this equation was derived by Lindblad \cite{Lindblad76}.

The same physical process of continuous measurement or gradual
decoherence may be presented selectively, i.e. with the readout of the
measurement (or the state of the environment) taken into account.
If some continuous measurement (continuous decoherence) of a system is
described selectively, the state of the system stays pure and may be
presented by a wave function (state vector) instead of a density
matrix.\footnote{Of course, this is valid only in the ideal case when
complete information about the measurement readout is known and
taken into account. If this information is partly unknown or partly
not taken into account, the state of the system is mixed, but the
degree of mixture is less than in the completely non-selective
description.}

The selective description of the process may be obtained without
making use of any explicit model of the environment if one apply the
approach based on restricted path integrals (RPI), see
\cite{Men00bk,Men93bk} and references therein. RPI approach to
continuous quantum measurement is thus model-independent. All that one
needs in this approach is the information obtained in the measurement
(recorded in the environment of the measured system). The concrete
realization of the environment and the form of recording the
information are irrelevant. This is an important feature of the
process in question, and this feature is quite clearly revealed by the
RPI approach.

If some observables are continuously measured, and resolution of time
in this measurement is considered to be absolute (the Markovian
approximation), then the evolution of the measured system is presented
by Schr\"odinger equation with a complex Hamiltonian. The transition
to the non-selective description gives then Lindblad equation. In the
non-Markovian approximation, when the resolution of time is assumed
finite, the evolution is readily presented by RPI but cannot be
reduced to a time-differential equation \cite{Men00bk,Men97timeRes}.

The description of continuous measurements (in other words, of slow
decoherence) by RPI was first derived \cite{Men79} in the
model-independent way from Feynman's path-integral form of quantum
mechanics. Later the restricted-path-integral description of some
processes of this type was derived from concrete models by
conventional quantum-mechanical methods \cite{Men00bk,Men97timeRes}.
In particular, an explicit model of a wide class of fuzzy continuous
measurements on a two-level system was considered in this way and the
Gaussian RPI (equivalently, quadratic imaginary potential) was derived
for the resulting evolution. This may be interpreted as a special case
of the quantum version of Central Limiting Theorem \cite{Men00bk}.

\quad

In the present paper the most general derivation of the RPI
description for the evolution of an open system is given.

For this end an arbitrary open system is considered and the evolution
operator is constructed for the larger system consisting of the system
of interest together with its environment. Then this evolution
operator is decomposed into the sum of partial evolution operators.
The condition is presented which provides that the partial evolutions
decohere. This condition is analogous to the consistency condition of
Gell-Mann and Hartle \cite{GellMannHartle93}. If this condition is
satisfied, probabilities may be correctly ascribed to the alternative
partial evolutions. In this case the partial evolutions may be
considered as classical alternatives.

Besides, a stronger so-called environment-decoherence condition is
introduced which means that decoherence of the alternatives follow
from the properties of the environment rather than of the system of
interest. The environment-decoherence condition is analyzed in terms
of the partial influence functionals (PIF) which provide a
decomposition of the Feynman-Vernon influence functional
\cite{FeynVernon63}.

One more condition on PIF is also considered which is formally more
strong but practically equivalent if the environment is mesoscopic or
macroscopic. This {\em condition for decoherence of PIF} is shown to
lead to the RPI description of the evolution of the open system. If
this condition is valid, the influence of the environment on the
system of interest may be interpreted as a continuous measurement of
this system by its environment.

It is argued that the choice of the set of decohering partial
evolutions (and therefore interpretation of the open system as a
continuously measured one) is possible if the environment is
macroscopic or mesoscopic.

\section{Evolution of an open system}

Let us consider an open system $\Sys$ which interacts with its
environment $\E$ according to the following scheme: \\

\centerline{%
\framebox{{\bf System}} $\leftrightarrow$ \framebox{{\bf Environment}}
}

\quad

\noindent
We shall
consider the evolution of the compound system $\Sys+\E$ and interpret
it in terms of alternative `partial evolutions'. Our goal will be to
formulate the conditions on the choice of the set of alternatives
which provide that the evolution may be interpreted as a continuous
measurement of the (open) system $\Sys$ by its environment $\E$, with
the given alternatives being the measurement readouts.

Note that the environment $\E$ may be a measuring device constructed
in a special way to perform a continuous measurement of $\Sys$, or a
`reservoir' interacting with the system of interest $\Sys$ in an
uncontrollable way. However even in the latter case
the process may be interpreted as a measurement.

\subsection{Evolution of the system and its environment}

The compound system consisting of $\Sys$ and $\E$ is assumed closed
(isolated). Let the Hamiltonian of this system have the form
\be
H_{\rm tot}(p,q,P,Q)=H(p,q)+H_{\E}(P,Q)+H_{I}(p,q,P,Q)
\label{TotalAction}\ee
where the Hamiltonians $H$, $H_{\E}$ and $H_{I}$ describe
correspondingly our system, its environment and their interaction (of
course, all the variables may be many-component, the Hamiltonians may
in principle depend on time).

The evolution of the compound system is given by the formula
\be
\R=U^{\rm tot}\, \R_{\rm in} \, ({U^{\rm tot}})^{\dag}
\label{TotEvol}\ee
where $\R$ is the density matrix of the compound system
at an arbitrary time moment, $\R_{\rm in}$
the density matrix at the initial time and $U^{\rm tot}$ the evolution
operator of the compound system during the corresponding time interval.

The evolution operator may be expressed in the form of the Feynman
path integral\footnote{We use here an integral over paths $[p,q,P,Q]$
in phase space of the compound system, correspondingly the action is
expressed in terms of the Hamiltonian of this system.}
\ba
\lefteqn{U^{\rm tot}=\int d[p,q] \int d[P,Q]}\nonumber\\
&\times&\exp \left\{
\ioverh \left[A[p,q]+A_{\E}[P,Q]-\int_0^t dt \, H_{I}(p,q,P,Q)\right]
\right\}
\label{TotEvolOper}\ea
where the following notations are introduced:
\be
A[p,q]=\int_0^t dt\, \left[p\dot{q} - H(p,q)\right], \quad
A_{\E}[P,Q]=\int_0^t dt\, \left[P\dot{Q} - H_{\E}(P,Q)\right] .
\label{Action}\ee

The evolution of the subsystem ${\cal S}$ may be presented by the
reduced density matrix $\rho$ obtained from $\R$ by tracing in the
degrees of freedom of the subsystem $\E$:
\be
\rho=\Tr_{\E} \R
= \Tr_{\E} \left[ U^{\rm tot}\, \R_{\rm in} \,
({U^{\rm tot}})^{\dag}\right] .
\label{ReducedDensMatrix}\ee

\subsection{Decomposition of the evolution operator}

We need a decomposition of the evolution operator $U^{\rm tot}$ in the
sum (integral) of {\em partial evolution operators} $U^{\rm
tot}_{\al}$
\be
U^{\rm tot}=\int  d\al \, U^{\rm tot}_{\al}
\label{DecomposEvol}\ee
(later on the alternatives $\al$ will be interpreted as
readouts of a continuous measurement).
The partial evolution operators may be defined as
\ba
\lefteqn{U^{\rm tot}_{\al}=\int d[p,q] \int d[P,Q] \;\;W_{\al}[P,Q] }
\label{PartEvolOper}\\
&\times& 
\exp \left\{
\ioverh \left[A[p,q]+A_{\E}[P,Q]-\int_0^t dt \, H_{I}(p,q,P,Q)\right]
\right\}
\nonumber\ea
where the {\em weight functionals}\footnote{Note that the weight
functionals may be chosen in principle in a more general form
$W_{\al}[p,q,P,Q]$ (for example in \Eqs{DecoFunc}{ConsistCond}), but
we need only the special case $W_{\al}[P,Q]$.} $W_{\al}[P,Q]$ form a
{\em decomposition of unity}, i.e. the relation
\be
\int  d\al \, W_{\al}[P,Q] = 1
\label{DecomposUnity}\ee
is valid with a certain measure $d\al$ on the set of all possible
alternatives $\al$.

Making use of the decomposition \eq{DecomposEvol} we can present the density
matrix $\R$ of the compound system as an integral of {\em partial
density matrices} expressing the {\em `partial evolutions'}:
\be
\R=\int  d\al d\beta \, \R_{\al\beta},
\qquad
\R_{\al\beta}=U^{\rm tot}_{\al}\, \R_{\rm in} \,
({U^{\rm tot}_{\beta}})^{\dag} .
\label{PartEvol}\ee

If the compound system's density matrix is decomposed as in
\Eq{PartEvol} then the corresponding decomposition of the reduced
density matrix of the subsystem $\Sys$ results:
\be
\rho=\int d\al d\beta \; \rho_{\al\beta},
\qquad
\rho_{\al\beta}=\Tr_{\E} \R_{\al\beta} .
\label{SysPartEvol}\ee

\section{Decoherence}

Let us now formulate the condition under which the alternatives
denoted as $\al$ are characterized by probabilities instead of
probability amplitudes so that there are no interference effects
between them. This is necessary for $\al$ to be interpreted as
classical alternatives (specifically, as measurement readouts of some
measurement).

\subsection{Decoherence of the system and its environment}

The (total) trace of the partial density matrices gives the set of
{\em generalized decoherence functionals}
\be
P_{\al\beta}=\Tr \R_{\al\beta},
\qquad
\int  d\al d\beta \, P_{\al\beta}=1 .
\label{DecoFunc}\ee
They are analogous to the decoherence functionals introduced by
Gell-Mann and Hartle \cite{GellMannHartle93} but for the alternatives
$\al$ presented by $W_{\al}$ (approximately presented by sets of paths
or `corridors of paths') instead of quantum histories (sequences of
projectors) used in \cite{GellMannHartle93}.

In analogy with the definition of Gell-Mann and Hartle, the
alternatives $\al$ are said to decohere if the following {\em
generalized consistency condition} is valid:
\be
P_{\al\beta}=\Tr \R_{\al\beta}=\delta(\al,\beta)\,P_{\al},
\qquad
\int  d\al \, P_{\al}=1 .
\label{ConsistCond}\ee
Here $\delta(\al,\beta)$ is a delta-function in respect to the
measure $d\al$ used in \eq{DecomposUnity}. \Eq{ConsistCond}
provides that the {\em alternatives $\{ \al \}$ decohere}, i.e. are
characterized by the {\em probabilities} $P_{\al}$ (more precisely, by
probability densities in respect to the measure $d\al$) rather than
probability amplitudes. This is necessary for the set $\{ \al \}$ to
be considered as the set of {\em classical alternatives}. Usually
condition \eq{ConsistCond} is satisfied only approximately.

\subsection{Decoherence induced by the environment}

Sometimes (typically in the situation of measurement) decoherence
may be a consequence of some properties of the environment $\E$, but
not of the system $\Sys$. Intuitively it is clear that this must take
place if the environment has great number of degrees of freedom,
especially if this number is macroscopic. However, instead of great
number of degrees of freedom this may be provided by special
properties of the environment (namely, by `orthogonality'
of the sets of paths corresponding to different alternatives).

The important situation when decoherence is caused only by the
environment may be mathematically formulated as follows:
delta-function $\delta(\al,\beta)$ arises not only after total tracing
as in \Eq{ConsistCond} but already after tracing in the environments's
degrees of freedom. This means that the relation
\be
\rho_{\al\beta}=\Tr_{\E} \R_{\al\beta}=\delta(\al,\beta)\,\rho_{\al}
\label{ConsCondEnv}\ee
should be valid with a set $\rho_{\al}$ of {\em `decohering partial
reduced density matrices'}. The total reduced density matrix will then
be decomposed as
\be
\rho=\int  d\al \, \rho_{\al} .
\label{EnvDecoDecompos}\ee

The condition \eq{ConsCondEnv} is stronger then the (generalized)
consistency condition \eq{ConsistCond}. It may be called the {\em
environment-decoherence condition}. One may expect that this condition
is approximately fulfilled if the environment contains {\em many
degrees of freedom} but not only in this case. Below we shall show
that a slightly stronger condition provides that the evolution of the
system $\Sys$ may be presented as a result of its continuous
measurement by the environment $\E$. The alternatives $\al$ may then
be interpreted as {\em measurement readouts} for this continuous
measurement.

\section{Partial influence functionals and RPI}\label{SectEnvDecoRPI}

Now we shall introduce Feynman-Vernon influence functional, decompose
it into partial influence functionals (PIF) and express the
environment-decoherence condition as well as some slightly stronger
condition in terms of PIF. This will allow us to go over to the
restricted-path-integral (RPI) description of the evolution of the
open system. The process may then be interpreted as a continuous
measurement.

\subsection{Partial influence functionals}

With \Eq{TotEvolOper} taken into account, the reduced density matrix
\eq{ReducedDensMatrix} may be expressed in the form of a multiple
functional integral. Assuming that the initial density matrix of the
compound system is factorized
\be
\R_{\rm in}(q,Q|\bar{q},\bar{Q})
=\R^{\E}_{\rm in}(Q,\bar{Q})\, \rho_{\rm in}(q,\bar{q})
\label{InDensMatrFactor}\ee
and performing the functional integrations in two
steps, we may put the reduced density matrix into the form
\ba
\lefteqn{\rho(q',\bar{q'})=
\int dq \int d\bar{q}
\int_{q}^{q'} d[p,q] \int_{\bar{q}}^{\bar{q'}} d[\bar{p},\bar{q}]}
\nonumber\\
&\times& \exp \left\{
\ioverh \left(A[p,q]-A[\bar{p},\bar{q}]\right)
\right\}
F[p,q|\bar{p},\bar{q}] \,
\rho_{\rm in}(q,\bar{q})
\label{DensMatrInflFunc}\ea
where
\ba
F[p,q|\bar{p},\bar{q}]
&=&\int dQ' \int dQ \int d\bar{Q}
\int_{Q}^{Q'} d[P,Q] \int_{\bar{Q}}^{Q'} d[\bar{P},\bar{Q}]
\;\R^{\E}_{\rm in}(Q,\bar{Q})\nonumber\\
&\times& \exp \left\{
\ioverh \left[A_{\E}[P,Q]-\int_0^t dt \, H_{I}(p,q,P,Q)\right]
\right\}
\nonumber\\
&\times& \exp \left\{
-\ioverh \left[A_{\E}[\bar{P},\bar{Q}]-\int_0^t dt \,
H_{I}(\bar{p},\bar{q},\bar{P},\bar{Q})\right]
\right\}
\label{InflFunc}\ea
is Feynman-Vernon's {\em influence functional} \cite{FeynVernon63}
in phase-space representation.

The partial reduced density matrices \eq{SysPartEvol} are then
expressed as
\ba
\lefteqn{\rho_{\al\beta}(q',\bar{q'})=
\int dq \int d\bar{q}
\int_{q}^{q'} d[p,q] \int_{\bar{q}}^{\bar{q'}} d[\bar{p},\bar{q}]}
\nonumber\\
&\times&\exp \left\{
\ioverh \left(A[p,q]-A[\bar{p},\bar{q}]\right)
\right\}
F_{\al\beta}[p,q|\bar{p},\bar{q}] \,
\rho_{\rm in}(q,\bar{q})
\label{PartDensMatrInflFunc}\ea
in terms of the {\em partial influence functionals}
\ba
F_{\al\beta}[p,q|\bar{p},\bar{q}]
&=&\int dQ' \int dQ \int d\bar{Q}
\int_{Q}^{Q'} d[P,Q] \int_{\bar{Q}}^{Q'} d[\bar{P},\bar{Q}]
\nonumber\\
&\times& W_{\al}[P,Q] \,
\R^{\E}_{\rm in}(Q,\bar{Q})\,
W^*_{\beta}[\bar{P},\bar{Q}]\, \nonumber\\
&\times& \exp \left\{
\ioverh \left[A_{\E}[P,Q]-\int_0^t dt \, H_{I}(p,q,P,Q)\right]
\right\}
\nonumber\\
&\times& \exp \left\{
-\ioverh \left[A_{\E}[\bar{P},\bar{Q}]-\int_0^t dt \,
H_{I}(\bar{p},\bar{q},\bar{P},\bar{Q})\right]
\right\}
\label{PartInflFunc}\ea
The complete influence functional $F$ is equal to the integral of
$F_{\al\beta}$ in analogy with \Eqs{PartEvol}{SysPartEvol}.

The environment-decoherence condition \eq{ConsCondEnv} means that the
partial influence functional $F_{\al\beta}$ is diagonal, i.e. contains
the delta-function $\delta(\al,\beta)$ as a factor. We shall argue
that under rather general assumptions the stronger condition
\be
F_{\al\beta}[p,q|\bar{p},\bar{q}]
=\delta(\al,\beta)\, w_{\al}[p,q]\, w^*_{\beta}[\bar{p},\bar{q}]
\label{DecoPartInflFunc}\ee
takes place. This condition may be called the {\em condition for
decoherence of the partial influence functionals}. It will be shown
(see Sect.~\ref{SectDerivRPI}) to lead to the RPI description of the
evolution of $\Sys$ so that the influence of the environment may be
considered as a continuous measurement of the system $\Sys$.

\subsection{Analysis of the condition for the measurement-type
evolution}\label{SectAnalisDecoCond}

The condition \eq{DecoPartInflFunc} turns out to be (approximately)
valid if the alternatives $\al$ (determined by the weight functionals
$W_{\al}$, see \Eq{PartEvolOper}) are chosen properly, so that the
corresponding sets (corridors) of paths $[P,Q]$ are sufficiently
wide, but not too wide.\footnote{We mean the sets of paths $[P,Q]$ in
phase space of the environment $\E$. The sets (corridors) of paths
$[p,q]$ in the phase space of the system $\Sys$ which are described by
the weight functionals $w_{\al}[p,q]$ may turn out to be arbitrarily
narrow.}

More precisely, the corridors (denote them $\al_{\E}$) must be chosen
in such a way that 1)~if a corridor contains no classical trajectory
(of the system $\E$), then the integral over this corridor is
negligibly small, and 2)~if a corridor contains some classical
trajectory then the integral over this corridor is well approximated
by the exponential of the action along this trajectory (the stationary
phase approximation).

If these conditions are satisfied, then $F_{\al\beta}$ is not
negligible in the sole case when both corridors $\al_{\E}$,
$\beta_{\E}$ contain classical trajectories and all these classical
trajectories are close to each other. Therefore, $F_{\al\beta}$ is
negligible each time when either one of $\al_{\E}$, $\beta_{\E}$
contains no classical trajectory or both of them contain classical
trajectories but those in $\al_{\E}$ strongly differ from those in
$\beta_{\E}$.\footnote{The classical trajectories contained in the
same corridor cannot strongly differ because of the assumed properties
of the corridors.}

These features are approximately presented by the formula
\eq{DecoPartInflFunc}. The functional $w_{\al}[p,q]$ in this formula
turns out to be negligible if the corresponding set of paths
$\al_{\E}$ of the environment (determined by the functional
$W_{\al}[P,Q]$) contains no classical trajectory.

\quad

Let us be somewhat more concrete and derive the concrete formulas for
those PIF which are not negligible, i.e. those $F_{\al\beta}$ that
both corridors $\al_{\E}$ and $\beta_{\E}$ (defined correspondingly by
$W_{\al}[P,Q]$ and $W_{\beta}[P,Q]$) contain classical trajectories.
The corridors were assumed to be sufficiently wide. This is why the
path integrals in \eq{PartInflFunc} may be approximately presented by
the exponentials of the classical action calculated along these
classical trajectories:
\ba
F_{\al\beta}[p,q|\bar{p},\bar{q}]
&\approx&\int_{I_{\rm fin}(\al)\cup I_{\rm fin}(\beta)} dQ'
\int_{I_{\rm in}(\al)} dQ \int_{I_{\rm in}(\beta)} d\bar{Q} \;\;
\R^{\E}_{\rm in}(Q,\bar{Q}) \nonumber\\
&\times&
\exp \left\{ \ioverh \left[
S_{\rm cl}(Q,Q',[p,q])
-S_{\rm cl}(\bar{Q},Q',[\bar{p},\bar{q}])
\right] \right\} .
\label{PartInflFuncApprox}\ea
Here $I_{\rm in}(\al)$ (correspondingly $I_{\rm fin}(\al)$) is the set
of initial (correspondingly final) points of those classical
trajectories which lie in the corridor $\al_{\E}$. The classical action
$S_{\rm cl}(Q,Q',[p,q])$ depends on the initial and final points $Q,Q'$
of the corresponding classical trajectory. Dependence on the path
$[p,q]$ arises because, according to \Eq{PartInflFunc}, this path
determines the force acting on the system $\E$ so that the shape of
the classical trajectory of this system depends on $[p,q]$. The same
is valid for $S_{\rm cl}(\bar{Q},Q',[\bar{p},\bar{q}]$.

The structure of the double path integral \eq{PartInflFunc} allows one
to consider it as a single path integral over trajectories `closed in
time' (going from the initial time moment to the time moment $t$ and
then backward to the initial time moment). In the part of the
integral, which corresponds to the inverse order of time, one has to
change the signs of the momentum $P$ and velocity $\dot{Q}$.

This is why the classical trajectory coming during the time interval
$[0,t]$ from the point $Q$ to the point $Q'$, should then, during the
interval $[t,0]$, return closely to the initial point. Therefore the
points $Q$ and $\bar{Q}$ should be close to each other.\footnote{They
precisely coincide if the paths $[p,q]$ and $[\bar{p},\bar{q}]$
coincide.} Approximately we may accept that $\bar{Q}=Q$.
\Eq{PartInflFuncApprox} takes then the form
\ba
F_{\al\beta}[p,q|\bar{p},\bar{q}]
&\sim&\int_{I_{\rm fin}(\al)\cup I_{\rm fin}(\beta)} dQ'
\int_{I_{\rm in}(\al)\cup I_{\rm in}(\beta)} dQ \;\;
\R^{\E}_{\rm in}(Q,Q) \nonumber\\
&\times&
\exp \left\{ \ioverh \left[
S_{\rm cl}(Q,Q',[p,q])
-S_{\rm cl}(Q,Q',[p,q])
\right] \right\} .
\label{PartInflFuncApprox2}\ea

\Eq{PartInflFuncApprox2} allows one to justify the approximate
expression \eq{DecoPartInflFunc} for PIF, particularly and clarify the
relations between the weight functionals $W_{\al}[P,Q]$ (or the
corresponding corridors of paths $\al_{\E}$) for the environment $\E$
and the functionals $w_{\al}[p,q]$ (and the corresponding corridors
which will be denoted by $\al$) for the system $\Sys$.

First of all, the integration regions $I_{\rm in}(\al)\cup I_{\rm
in}(\beta)$ and $I_{\rm fin}(\al)\cup I_{\rm fin}(\beta)$ in
\Eq{PartInflFuncApprox2} show that the integral is non-zero only if
the alternatives $\al$ and $\beta$ are very close to each other. This
is approximately presented by the delta-function $\delta(\al,\beta)$.

Consider now the dependence of the classical trajectories, along which
the classical actions in \Eq{PartInflFuncApprox2} are calculated,
correspondingly on the paths $[p,q]$, $[\bar{p},\bar{q}]$. The
expression \Eq{PartInflFuncApprox2} is written under the assumption
that the paths $[p,q]$, $[\bar{p},\bar{q}]$ are such that both
classical trajectories are close to the middles of the corridors
$\al_{\E}$, $\beta_{\E}$.

Denote by $[p_{\al},q_{\al}]$ that path of $\Sys$ which provides the
classical trajectory of the system $\E$ to be in the middle of the
corridor $\al_{\E}$. If the alternatives $\al$, $\beta$ coincide with
each other and both $[p,q]$, $[\bar{p},\bar{q}]$ coincide with
$[p_{\al},q_{\al}]$, then the expression \eq{PartInflFuncApprox2} is
real and has maximum absolute value. If $[p,q]$, $[\bar{p},\bar{q}]$
deflect from $[p_{\al},q_{\al}]$, then the expression
\eq{PartInflFuncApprox2} is not valid. Instead, \Eq{DecoPartInflFunc}
arises. The factors $w_{\al}[p,q]$, $w_{\beta}[\bar{p},\bar{q}]$
become smaller and smaller in absolute value with the deflections
increasing, and may become complex. The factor $w_{\al}[p,q]$
(correspondingly $w_{\beta}[\bar{p},\bar{q}]$) become negligible when
$[p,q]$ (correspondingly $[\bar{p},\bar{q}]$) is such that the
corridor $\al_{\E}$ (correspondingly $\beta_{\E}$) does not contain
classical trajectories.

\subsection{Derivation of RPI}\label{SectDerivRPI}

Let the alternatives $\al$ be chosen properly (as it is discussed in
\Sect{SectAnalisDecoCond}) so that \Eq{DecoPartInflFunc} is valid. In
this case the alternatives $\al$ decohere, i.e. can be considered as
classical alternatives. It may be readily shown that in this case the
system $\Sys$ evolves just as it is suggested in the
restricted-path-integral (RPI) description of continuous measurements.
The alternatives $\al$ play then the role of {measurement readouts}.

Indeed, if the condition \eq{DecoPartInflFunc} is valid,
then,according to \eq{PartDensMatrInflFunc}, the partial density
matrices of the system ${\cal S}$ take the form
\be
\rho_{\al\beta}
=\delta(\al,\beta)\; U_{\al}\,\rho_{\rm in} \, U_{\al}^{\dag}
\label{DecoEvolRho}\ee
where the following {\em partial evolution operator for the open
system $\Sys$} arises:\footnote{It should not be confused with the
partial evolution operator $U^{\rm tot}_{\al}$ of the closed compound
system $\Sys+\E$.}
\be
U_{\al}
=\int d[p,q]\, w_{\al}[p,q]\,
\exp \left\{
\ioverh \int_0^t \left[p\dot{q} - H(p,q)\right]
\right\} .
\label{RPI}\ee
This leads to the following evolution law for the open
system $\Sys$:
\be
\rho_{\al} = U_{\al}\, \rho_{\rm in} \, U^{\dag}_{\al}, \quad
\rho = \int d\al \, \rho_{\al}
= \int d\al \, U_{\al}\, \rho_{\rm in} \, U^{\dag}_{\al} .
\label{DecoEvolRho2}\ee

The first formula in \eq{DecoEvolRho2} presents a selective
description of the evolution of $\Sys$ (the evolution conditioned by
the alternative $\al$) while the second formula gives the
non-selective description of the same process (all possible
alternatives are taken into account).

The selective form of the evolution law may equivalently be expressed
in terms of the state vectors instead of density matrices:
\be
|\psi_{\al}\ra
=U_{\al}|\psi_{\rm in}\ra .
\label{DecoEvolPsi}\ee

The formulas \eqss{DecoEvolRho2}{DecoEvolPsi}{RPI} are characteristic
for the RPI approach to description of a continuous measurement
\cite{Men00bk}. The alternatives $\al$ are interpreted in this
approach as alternative readouts of the continuous measurement.
Therefore, we derived the RPI description of a continuous measurement
starting from the standard description of an open system and choosing
the set of alternatives $\al$ in such a way that the
environment-decoherence condition \eq{ConsCondEnv} and the close
condition \eq{DecoPartInflFunc} on the partial influence functionals
are satisfy.

This gives an one more justification of RPI description of
continuously measured (open, gradually decohering) systems. Moreover,
this shows that a wide class of open systems may be interpreted as
systems continuously measured by their environments.

\section{Conclusion}

In the previous papers the restricted-path-integral (RPI)
description of continuous quantum measurements was derived directly
from the Feynman's path-integral form of quantum mechanics. Besides,
the RPI behavior of measured systems was derived also by conventional
quantum-mechanical methods in the framework of some models of
measurements.

Here we considered an open system and decomposed its evolution into
the sum of partial evolutions. It was showed that the evolution of
the open system may be correctly described in the framework of the RPI
approach provided that the set of alternative partial evolutions
decohere as a consequence of the specific features of the environment.

Mathematically the condition for the measurement-type evolution of an
open system may be formulated as the {\em condition for decoherence of
the partial influence functionals} \eq{DecoPartInflFunc} or with the
help of the weaker (but practically equivalent for large environments)
{\em environment-decoherence condition} \eq{ConsCondEnv}.

These conditions are (approximately) valid if the alternatives $\al$
are presented by such sets of paths (corridors of paths) of the
environment which possess classical properties. The choice of
decohering alternatives is possible if the environment have many
degrees of freedom (is macroscopic or mesoscopic).

The results of the above consideration give a one more justification
of the restricted-path-integral approach to quantum continuous
measurements. This confirms that Feynman's theory of amplitudes and
Feynman's path integral technics are valid not only for closed
systems, but also in case of open systems. This is important because
Feynman's formalism provides a physically transparent presentation of
quantum mechanics and is efficient both for calculation and for
euristic considerations.

\quad

\noindent
\textbf{Acknowledgement.}  This work is supported in part by the
Russian Foundation for Basic Research under grant 02-01-00534.

\end{document}